\documentclass[aps,showpacs,preprint,superscriptaddress]{revtex4}
\usepackage{graphicx}
\usepackage{subfigure}
\usepackage{float}
\usepackage{amsmath}
\usepackage{amsfonts}
\usepackage{color}
\usepackage{bm}
\usepackage{bbm}
\usepackage{txfonts}
\usepackage{array}
\usepackage{amssymb}

\begin{document}
\title{Enhancement of electron-positron pairs in combined potential wells with linear chirp frequency}
\author{Li Wang}
\affiliation{Key Laboratory of Beam Technology of Ministry of Education, Beijing Radiation Center, Beijing 100875, China}
\affiliation{Key Laboratory of Beam Technology of Ministry of Education,
College of Nuclear Science and Technology, Beijing Normal University, Beijing 100875, China}

\author{Lie-Juan Li}
\affiliation{Key Laboratory of Beam Technology of Ministry of Education,
College of Nuclear Science and Technology, Beijing Normal University, Beijing 100875, China}

\author{Melike Mohamedsedik}
\affiliation{Key Laboratory of Beam Technology of Ministry of Education,
College of Nuclear Science and Technology, Beijing Normal University, Beijing 100875, China}

\author{Rong An}
\affiliation{Key Laboratory of Beam Technology of Ministry of Education, Beijing Radiation Center, Beijing 100875, China}
\affiliation{Key Laboratory of Beam Technology of Ministry of Education,
College of Nuclear Science and Technology, Beijing Normal University, Beijing 100875, China}

\author{Jing-Jing Li}
\affiliation{Key Laboratory of Beam Technology of Ministry of Education, Beijing Radiation Center, Beijing 100875, China}
\affiliation{Key Laboratory of Beam Technology of Ministry of Education,
College of Nuclear Science and Technology, Beijing Normal University, Beijing 100875, China}

\author{Bai-Song Xie}
\email[Corresponding author. Email: ]{bsxie@bnu.edu.cn}
\affiliation{Key Laboratory of Beam Technology of Ministry of Education, Beijing Radiation Center, Beijing 100875, China}
\affiliation{Key Laboratory of Beam Technology of Ministry of Education,
College of Nuclear Science and Technology, Beijing Normal University, Beijing 100875, China}

\author{Feng-Shou Zhang}
\email[Corresponding author. Email: ]{fszhang@bnu.edu.cn}
\affiliation{Key Laboratory of Beam Technology of Ministry of Education, Beijing Radiation Center, Beijing 100875, China}
\affiliation{Key Laboratory of Beam Technology of Ministry of Education,
College of Nuclear Science and Technology, Beijing Normal University, Beijing 100875, China}

\begin{abstract}
The effect of linear chirp frequency on the process of electron-positron pairs production from vacuum in the combined potential wells is investigated by computational quantum field theory. Numerical results of electron number and energy spectrum under different frequency modulation parameters are obtained. By comparing with the fixed frequency, it is found that frequency modulation has a significant enhancement effect on the number of electrons. Especially when the frequency is small, appropriate frequency modulation enhances multiphoton processes in pair creation, thus promoting the pair creation. However, the number of electrons created by high frequency oscillating combined potential wells decreases after frequency modulation due to the phenomenon of high frequency suppression. The contours of the number of electrons varying with frequency and frequency modulation parameters are given, which may provide theoretical reference for possible experiments.
\end{abstract}
\maketitle

\section{Introduction}
In recent decades, the research of electron-positron pairs created from vacuum under extreme external field is one of the hot topics.
The existence of positron was first predicted theoretically by Dirac in 1928\cite{Dirac1928}. Soon after that, it was confirmed in the laboratory by Anderson \cite{Anderson1933}. After pioneering researches of Sauter\cite{Sauter1931}, Heisenberg and Euler\cite{Heisenberg1936}, and Schwinger\cite{Schwinger1951}, several research methods were developed, such as proper time technique\cite{Tsai1973,Baier1974}, Wentzel-Kramers-Brillouin (WKB) approximation\cite{Kim2002}, worldline instanton technique\cite{Dunne2005,Dunne2006}, quantum kinetic method\cite{Alkofer2001,Roberts2002,Kluger1998,Hebenstreit2011}, and the computational quantum field theory\cite{Cheng2009,Xie2017,Lv2018}.

In the constant electric field $E$, pair creation probability is proportional to
$\exp(-\pi E_{\rm{cr}}/E)$. The Schwinger critical field strength is $E_{\rm{cr}}=10^{16}\rm{V/cm}$, which is achieved at a laser intensity of roughly $10^{29}\rm{W/cm}^2$. Since the required field strength is far above the presently achieved laboratory electric field strength, this prediction has not been directly verified experimentally.
With the rapid development of laser technique, it is believed that the laser intensity will be close to the Schwinger critical field strength in the near future.

At present, there are three mechanisms to create electron-positron pairs in vacuum under strong fields in theory. Electron-positron pairs are created by the Schwinger mechanism, which requires a strong field\cite{Jiang2011,Lv2013,Liu2015}.
In the multiphoton processes, the creation of electron-positron pairs under the alternating field requires a high oscillation frequency\cite{Jiang2012,Jiang2013,Tang2013,Hubbell2006}. By combining a strong and constant field with a weaker and high frequency oscillating field, electron-positrons pairs can be created through the dynamically assisted Sauter-Schwinger effect, where the pair-creation probability is strongly enhanced \cite{Schutzhold2008,Orthaber2011,Otto2015a,Otto2015b,Linder2015,Schneider2016,Torgrimsson2017,Torgrimsson2018,Sitiwaldi2018}.

In order to lower the threshold of pair creation and increase yield, many facilities were proposed, such as multiple well-barrier structures\cite{Gong2018}, the symmetric potential well\cite{Tang2013}, oscillating Sauter potential\cite{Wangli2019}, and asymmetric Sauter potential well\cite{Sawut2021}. The number of created electrons in the symmetric potential is more than that in the asymmetric potential due to the two edges of the potential well\cite{Tang2013}. In a static sauter potential well, although the bound state promotes the generation of electron-positron pairs, the inhibitory effect of Pauli blocking cannot be ignored\cite{Wangli2019,Su2020}.
In an oscillating potential well, Pauli blocking effect is weakened, but the multiphoton processes are highly depends on the oscillating frequency\cite{Jiang2013}.
By combining a static potential well with an alternating one, the bound state provides a bridge for the multiphoton process, and reduces the required frequency\cite{Tang2013}. According to energy spectrum analysis, the energy of created electron is affected by the energy level of bound state and laser frequency.
In the laboratory laser field, frequency modulation has more real significance because of chirp pulse amplification technique\cite{Strickland1985}.
Therefore, the influence of frequency modulation on electron-positron pairs in alternating fields attracted much attention\cite{Dumlu2010,Olugh2019,Abdukerim2017}.
In frequency modulated laser fields, the number of created electrons can be increased significantly for certain modulation parameters\cite{Gong2020,Liliejuan2021,Mohamedsedik2021}.
Enhanced electron-positron pairs production by frequency modulation in one- and two-color laser pulse fields is also studied\cite{Abdukerim2017}.
To our best knowledge, the effect of frequency modulation on pair creation under combined potential wells of a static potential well and an alternating one has not been studied.

In this paper, the enhancement of electron-positron pairs in the combined potential wells with linear chirp frequency is investigated by using the computational quantum field theory. First, the number of created electrons varying with chirp parameter is obtained. Then, the creation mechanism of electron-positron pairs is studied by analyzing the energy spectrum of created electrons. Furthermore, we show contour plot of the electron number varying with chirp parameters.

This paper is organized as follows. In Sec. II, we describe the outline of computational quantum field theory and illustrate combined potential wells.
In Sec.III, the number and energy spectrum of created electrons under different frequency-modulation parameters are simulated. The results under different parameters were compared and analyzed. In Sec.IV, we summarize our work.

\section{The framework of computational quantum field theory and the external field}\label{section2}
The computational quantum field theory is a physical theory that combines quantum mechanics with classical field theory. It provides an effective framework for describing multi-particle systems, especially those involving particle generation and annihilation processes.
The time evolution of electron-positron field operator $\hat{\psi}\left(z,t\right)$ can be obtained by Dirac equation\cite{Cheng2009}
\begin{equation}\label{Eq Dirac}
i\partial \hat{\psi} \left(z,t\right) / \partial{t}=\left[c\alpha_z \hat{P}+\beta c^2+V\left(z,t\right)\right] \hat{\psi}\left(z,t\right),
\end{equation}
where $\alpha_z$ and $\beta$ are the $z$ component and diagonal parts of the spin Dirac matrices, respectively, $c$ is the speed of light in vacuum, $V\left(z,t\right)$ is external field that varies with time $t$ in the $z$ direction. The atomic units $(\hbar=e=m_e=1)$ are used for convenience.
By introducing the creation and annihilation operators, the field operator $\hat{\psi}(z,t)$ can be decomposed as follows:
\begin{equation}\label{Eq Field Operator}
\begin{aligned}
\hat{\psi}(z,t)&=\sum_{p}\hat{b}_p(t)W_p(z)+\sum_{n}\hat{d}_n^{\dag}(t)W_n(z) \\
&=\sum_{p}\hat{b}_pW_p(z,t)+\sum_n\hat{d}_n^\dag W_n(z,t),
\end{aligned}
\end{equation}
where $\sum_{p(n)}$ represents summation over all positive$($negative$)$ energy, $p$ and $n$ denote the momenta of positive- and negative-energy states, $W_{p}(z)=\langle z|p\rangle$ and $W_{n}(z)=\langle z|n\rangle$ are positive and negative energy eigenfunctions of the field-free Dirac equation,
$W_{p}(z,t)=\langle z|p(t)\rangle$ and $W_{n}(z,t)=\langle z|n(t)\rangle$ are the time evolution of $W_{p}(z)$ and $W_{n}(z)$, respectively.

From Eq.(\ref{Eq Field Operator}), we obtain
\begin{equation}\label{Eq fermion operators}
\begin{aligned}
\hat{b}_p(t)&=\sum_{p'}\hat{b}_{p'}U_{pp'}(z,t)+\sum_{n'}\hat{d}_{n'}^\dag U_{pn'}(z,t),\\
\hat{d}_n^\dag(t)&=\sum_{p'}\hat{b}_{p'}U_{np'}(z,t)+\sum_{n'}\hat{d}_{n'}^\dag U_{nn'}(z,t),\\
\hat{b}_p^\dag(t)&=\sum_{p'}\hat{b}_{p'}^\dag U_{pp'}^*(z,t)+\sum_{n'}\hat{d}_{n'}U_{pn'}^*(z,t),\\
\hat{d}_n(t)&=\sum_{p'}\hat{b}_{p'}^\dag U_{np'}^*(z,t)+\sum_{n'}\hat{d}_{n'}U_{nn'}^*(z,t),
\end{aligned}
\end{equation}
where $U_{pp'}(t)=\langle p|\hat{U}(t)|p'\rangle$, $U_{pn'}(t)=\langle p|\hat{U}(t)|n'\rangle$, $U_{nn'}(t)=\langle n|\hat{U}(t)|n'\rangle$, and $U_{np'}(t)=\langle n|\hat{U}(t)|p'\rangle$, the time-ordered propagator $\hat{U}(t)=\textrm{exp}\{{-i\int^t d\tau [c\alpha_z \hat{p}+\beta c^2+V(z,\tau)]}\}$.

In Eq.(\ref{Eq Field Operator}), the electronic portion of the field operator is defined as $\hat{\psi}_e(z,t)\equiv \sum_p \hat{b}_p(t)W_p(z)$. So we can obtain the probability density of created electrons by
\begin{equation}\label{Eq Density}
\begin{aligned}
\rho(z,t)&=\langle \mathrm{vac}|\hat{\psi}_e^\dag (z,t)\hat{\psi}_e(z,t)|\mathrm{vac}\rangle \\
&=\sum_n |\sum_p U_{pn}(t)W_p(r)|^2
\end{aligned}
\end{equation}
By integrating this expression over all space, the number of created electrons can be obtained as
\begin{equation}\label{Eq Number}
N(t)=\int \rho(z,t)dz=\sum_p \sum_n |U_{pn}(t)|^2.
\end{equation}
The time-ordered propagator $U_{pn}(t)$ can be numerically calculated by employing the split-operator technique. Therefore, according to Eqs.(\ref{Eq Density}) and (\ref{Eq Number}) we can compute various properties of the electrons produced under the action of the external potential.

In this paper, our configuration of potentials is
\begin{equation}\label{Eq Well}
\begin{aligned}
\begin{split}
V(z,t)= \left \{
\begin{array}{ll}
S(z)[V_1+V_2]\cos(\pi t/2t_0),                        & 0\leq t<t_0\\
S(z)[V_1+V_2\cos(b(t-t_0)^2+\omega_0 (t-t_0)+\phi)],     & t_0\leq t<t_0+t_1\\
S(z)[V_1+V_2\cos(bt_1^2+\omega_0 t_1+\phi)]\cos[\pi(t-t_1-t_0)/2t_0] ,                                 & t_0+t_1\leq t\leq t_1+2t_0
\end{array}
\right.
\end{split}
\end{aligned}
\end{equation}
where $S(z)=\{\tanh[(z-D/2)/W]-\tanh[(z+D/2)/W]\}/2$, $D$ is the width of the potential well, and $W$ is the width of the potential well edge, which corresponds to the intensity of the electric field. $V_1$ and $V_2$ are the depth of the static and the oscillating potential well, respectively.
In cases $t_0<t<t_0+t_1$, combined potential wells consist of a static potential well with a depth of $V_1$ and an oscillating potential well with an amplitude of $V_2$. The oscillation frequency is linearly time dependent, where $\omega_0$ is the fundamental frequency, $b$ is the chirp parameter, describing the change in frequency over time.
In cases $0<t<t_0$ and $t_0+t_1<t<t_1+2t_0$, it describes turning on and off processes of combined potential wells, respectively.
For simplicity, the phase $\phi$ is set to $0$ throughout the paper.

\begin{figure}
\centering\includegraphics[width=10cm]{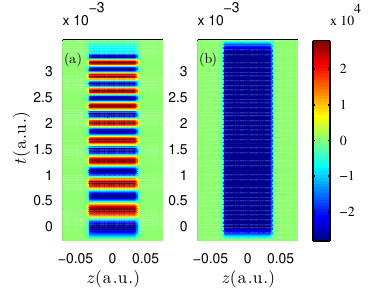}
\caption{\label{fig1} Contour profile plot of the space-time structure of combined potential wells. Panel (a) is for the oscillating potential well, and panel (b) is for the static potential well. The fundamental frequency is set to $\omega_0=0.5c^2$. The chirp parameter is set to $b=0.42c^2/t_1$, where $t_1=20\pi/c^2$a.u.. The space size is set to $L=2$. Other parameters are set to $V_1=V_2=1.5c^2$, $W=0.3\lambda_e$, $D=10\lambda_e$, $\lambda_e=1/c$, $t_0=5/c^2$a.u..}
\end{figure}

In Fig. \ref{fig1}, the contour profile plot of the space-time structure of combined wells is presented, where panel (a) and (b) are for the oscillating and the static potential well, respectively. The edge and width of combined potential wells are all set to $W=0.3\lambda_e$, $D=10\lambda_e$, where $\lambda_e=1/c$ is the Compton wavelength of the electron. The potential well depths are set to $V_1=V_2=1.5c^2$. The frequency parameters are set to $\omega_0=0.5c^2$, $b=0.42c^2/t_1$, where $t_1=20\pi/c^2$a.u. is the interaction time. The opening and closing time are all set to $t_0=5/c^2$a.u.. The unit of chirp parameter $b$ is set as $c^2/t_1$ here to facilitate the determination of the frequency at the last moment. For example, the effective frequency reaches $\omega_{eff}=0.92c^2$ at $t=t_1$ in Fig. \ref{fig1}. The space size is set to $L=2$, which ensures that all electrons generated during the numerical simulation are counted.

First, in Fig. \ref{fig1}(a), the depth of the oscillating potential well slowly reaches its maximum at $t=0$, and then the oscillation speed becomes faster and faster as time goes on.
During the closing process of oscillating potential well, the depth gradually approaches to $0$. Note that the depth is continuously and slowly varying from the value at $t=t_1$, rather than suddenly returning to the maximum. This reduces the error caused by the instantaneous opening and closing of the potential well.
In Fig. \ref{fig1}(b), the static potential well opens and closes slowly. The depth and the width remain unchanged during $0<t<t_1$, which provides stable bound states for the multiphoton processes.

\section{Numerical results}
In this section, numerical results are mainly presented in figures, where relevant parameters are set to $D=10\lambda_e$, $W=0.3\lambda_e$, $V_1=V_2=1.5c^2$, $t_0=5/c^2$a.u., $L=2$, $t_1=20\pi/c^2$a.u..

\begin{figure}[htbp]\suppressfloats
\centering\includegraphics[width=10cm]{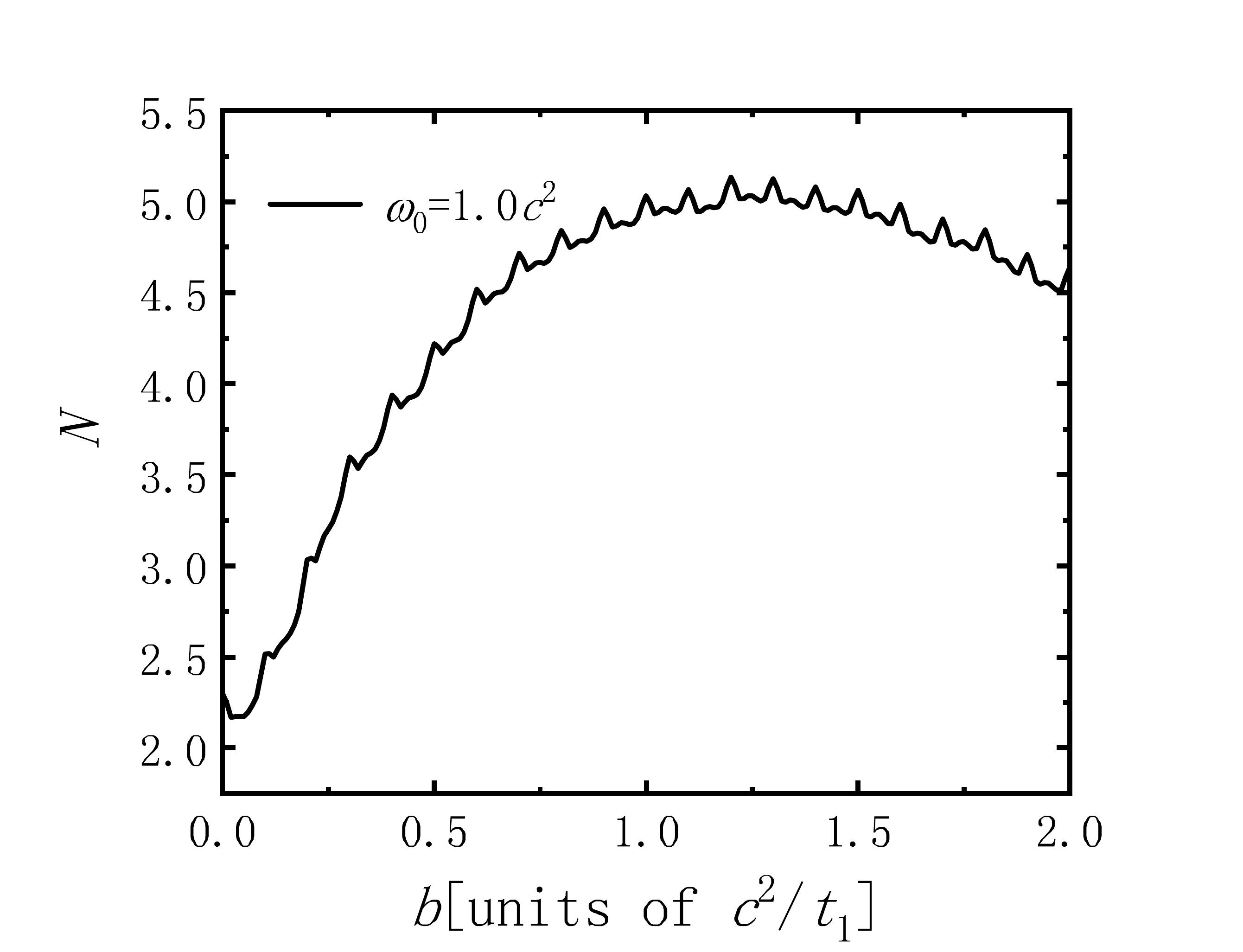}
\caption{\label{fig2} Number of created electrons as a function of chirp parameter $b$ for $\omega_0=1.0c^2$. Other parameters are the same as in Fig. \ref{fig1}. }
\end{figure}
The final number $N$ of created electrons varying with chirp parameter $b$ is presented in Fig. \ref{fig2}. The fundamental frequency is set to $\omega_0=1.0c^2$. In order to facilitate data analysis, the chirp parameter $b$ is set in the unit of $c^2/t_1$. When $b=0$, the frequency is fixed at $\omega_0=1.0c^2$. With the increase of chirp parameter $b$, the final number first increases quickly, then decreases slowly. When $b=1.2c^2/t_1$, the final number of created electrons reaches the maximum.
The frequency modulation has a significant enhancement effect on the number of electrons.
In addition, there are many peaks appear in Fig. \ref{fig2} with a period of $0.1c^2/t_1$. The oscillation phenomenon is caused by the time crystals made of electron-positron pairs, which describes the appearance and disappearance of electron-positron pairs\cite{Bialynicki-Birula2021}. The period of oscillation can be traced back to Eq. (\ref{Eq Well}). When the total depth of combined potential wells at $t=t_0+t_1$ reaches the maximum, i.e., $\omega_0 t_1+bt_1^2=2n\pi, n=1,2,3\ldots$, the generation rate of electrons increases to the maximum.
Therefore, when the chirp parameter $b$ increases by $2\pi/t_1^{2}$, or the fundamental frequency $\omega_0$ increases by $2\pi/t_1$, an additional peak occurs, which is verified in Figs. \ref{fig2}, \ref{fig5}, and \ref{fig6}. The peak period depends on the numerical simulation time $t_1$. Since the numerical simulation time cannot be infinite, the resulting error is inevitable, but it can be reduced by increasing $t_1$.

\begin{figure}[htbp]\suppressfloats
\centering\includegraphics[width=12cm]{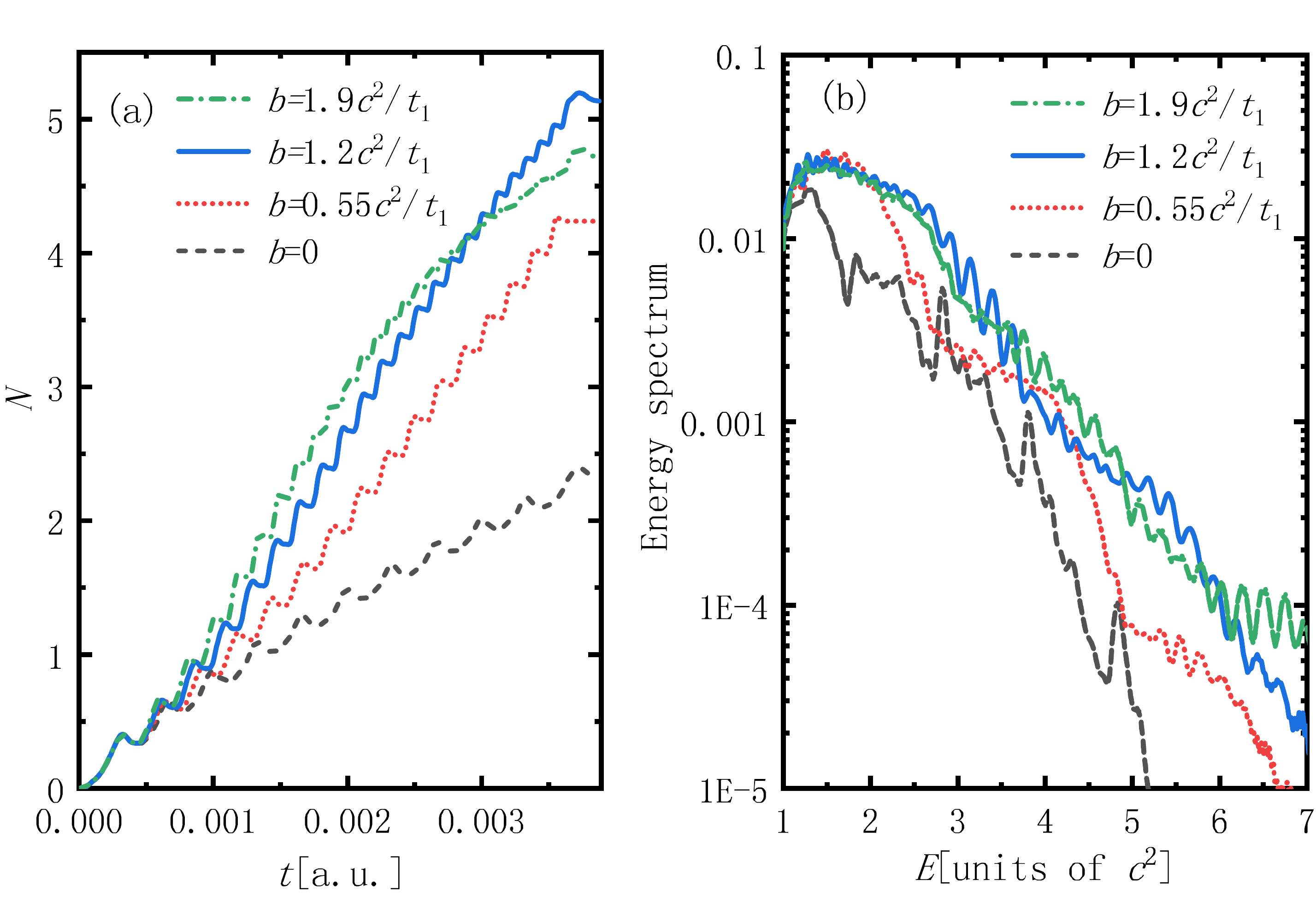}
\caption{\label{fig3} (a) Number of created electrons as a function of time and (b) Energy spectrum of created electrons for $b=0$ (the black dashed curve), $0.55c^2/t_1$ (the red dotted curve), $1.2c^2/t_1$ (the blue solid curve), and $1.9c^2/t_1$ (the green dotted and dashed curve). The fundamental frequency is set to $\omega_0=1.0c^2$. Other parameters are the same as in Fig. \ref{fig1}. }
\end{figure}
The electron number varying with time and energy spectrum of electrons are presented in Figs.\ref{fig3} (a) and (b), respectively. The fundamental frequency $\omega_0$ is the same as in Fig. \ref{fig2}. In Fig. \ref{fig3}(a), when $b=0$, the number of electrons increases with a fixed period over time. Instead, the period of the number of electrons generated under chirp frequency changing with time is getting shorter and shorter. This is due to a linear increase in frequency.
The final numbers for different chirp parameters are all consistent with Fig. \ref{fig2}. For $b=1.9c^2/t_1$, when $t>0.002$a.u., the pair production rate deceases significantly. Hence, the final number is less than that of $b=1.2c^2/t_1$, which is due to high frequency suppression.

\begin{figure}[htbp]\suppressfloats
\includegraphics[width=15cm]{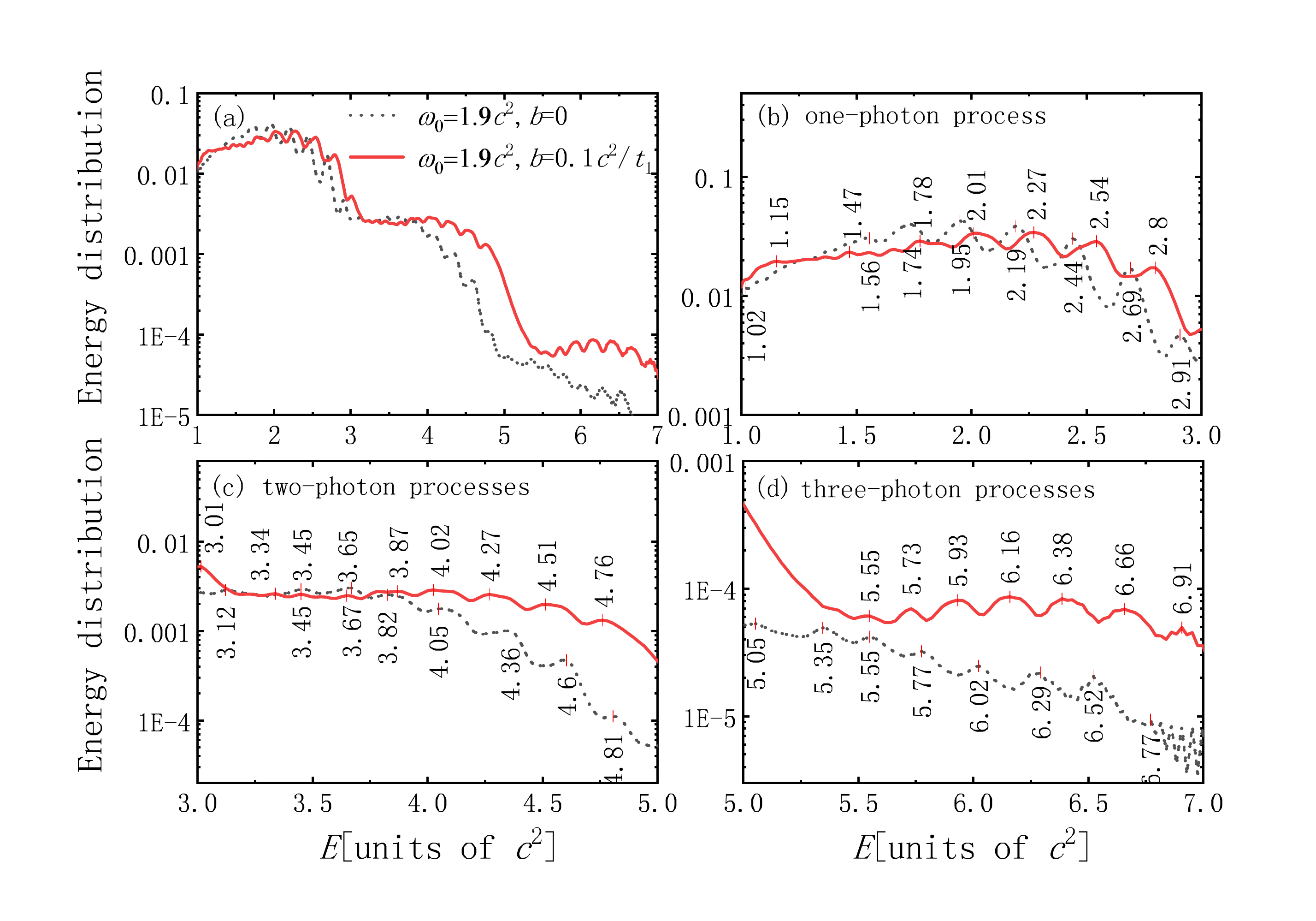}
\caption{\label{fig4} Energy spectrum of created electrons under the fixed frequency (the black dotted curve) with $\omega_0=1.9c^2$, $b=0$ and the chirp frequency (the red solid curve) with $\omega_0=1.9c^2$, $b=0.1c^2/t_1$. Other parameters are the same with Fig. \ref{fig1}.}
\end{figure}

\begin{figure}[htbp]\suppressfloats
\centering\includegraphics[width=10cm]{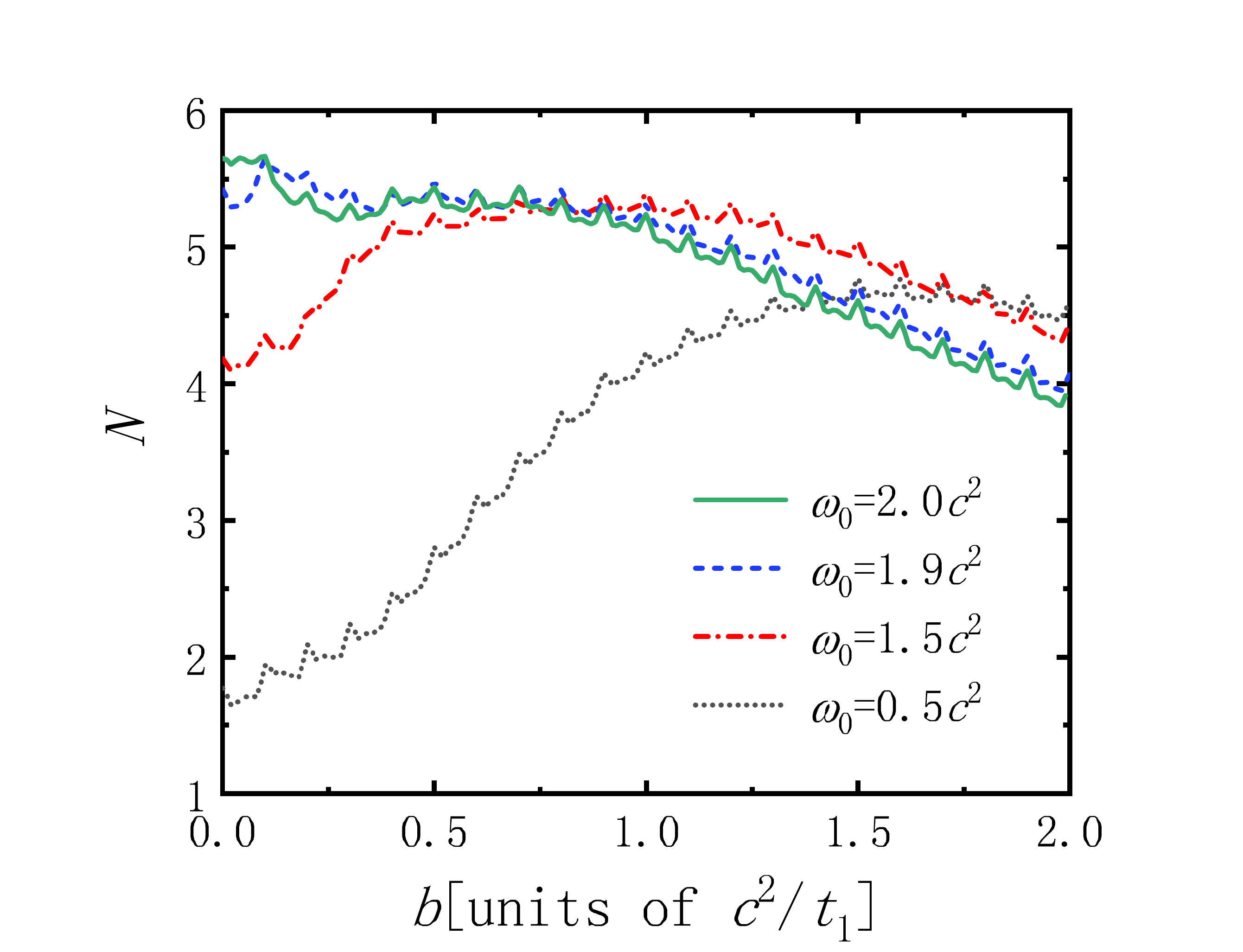}
\caption{\label{fig5} Number of created electrons as a function of the chirp parameter $b$ for $\omega_0=0.5c^2$ (the black dotted curve), $1.5c^2$ (the red dotted and dashed curve), $1.9c^2$ (the blue dashed curve), and $2.0c^2$ (the green solid curve). Other parameters are the same as in Fig. \ref{fig1}. }
\end{figure}

\begin{figure}[htbp]\suppressfloats
\includegraphics[width=10cm]{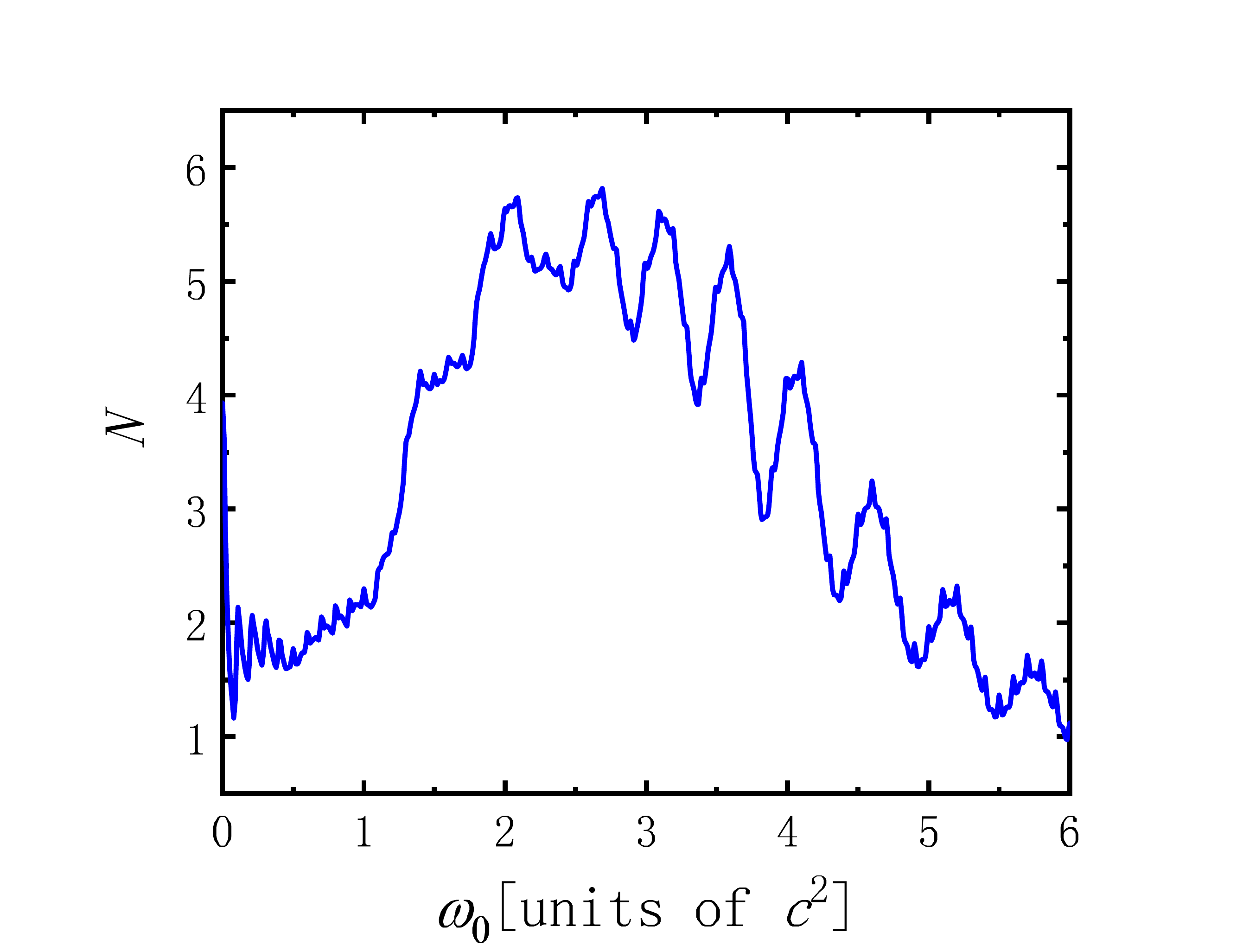}
\caption{\label{fig6} Number of created electrons varying with the fixed frequency $\omega_0$ for $b=0$. Other parameters are the same with Fig. \ref{fig1}.}
\end{figure}

In Fig. \ref{fig3}(b), the energy spectrum under fixed frequency has several distinct peaks and the narrow energy region. With the increase of chirp parameter $b$, the energy region widens and the number of peaks increases. This reflects that frequency modulation affects the multiphoton processes. The energy of photon absorbed through the multiphoton processes under a fixed frequency is single, so the peak number of energy spectrum is less and discrete. With the increase of time, frequency modulation widens the spectrum, which provides an abundance of photons of different energies. The electron energy spectrum becomes wider and the peak number increases, see Fig. \ref{fig4} for details.

In Fig. \ref{fig4}, the energy spectrum of created electrons under a fixed frequency and the chirp frequency is presented. The black dotted curve corresponds to the fixed frequency with $\omega_0=1.9c^2$, $b=0$. The solid red line is the chirp frequency with $\omega_0=1.9c^2$, $b=0.1c^2/t_1$. In Fig. \ref{fig4}(a), we present the whole energy spectrum of one-photon process, two-photon processes, and three-photon processes.
Compared with the fixed frequency, the energy spectrum under chirp frequency is wider.
In Figs. \ref{fig4}(b), (c), and (d), the influence of frequency modulation on energy conversion from photon to electrons in multiphoton processes can be analyzed in detail.
According to Ref. \cite{Tang2013}, when $D=10\lambda_e$, and $W=0.3\lambda_e$, combined potential wells have 8 bound-state levels, which are $E_1=-0.4247c^2$, $E_2=-0.3069c^2$, $E_3=-0.1361c^2$, $E_4=0.0680c^2$, $E_5=0.2919c^2$, $E_6=0.5260c^2$, $E_7=0.7618c^2$, $E_8=0.9778c^2$.
In Fig. \ref{fig4}(b), when the frequency is fixed, energy peaks $E=1.56c^2, 1.74c^2, 1.95c^2, 2.19c^2, 2.44c^2, 2.69c^2, 2.91c^2$ correspond to $E_{2}\sim E_{8}$ with the relation of $E=E_i+1.9c^2$ for $i=2$ to 8.
From the red solid curve, energy peaks $E'=2.01c^2, 2.27c^2, 2.54c^2, 2.8c^2$ are reduced for the relation of $E'=E_i+2.0c^2$ for $i=4$ to 7, energy peak $E'=1.47c^2$ is in the relation of $E'=E_1+1.9c^2$.
Obviously, the photon absorbed in multiphoton process under chirp frequency is not a single energy. With the increase of frequency, the photon energy increases from $1.9c^2$ to $2.0c^2$. Correspondingly, the energy spectrum shows that the energy of the absorbed photon ranges from $1.9c^2$ to $2.0c^2$.

In Fig. \ref{fig4}(c), energy peaks are in the relation of $E=E_i+2\omega_0$ for the fixed frequency, in relations of $E'=E_i+2(\omega_0+bt_1)$ and $E'=E_i+2\omega_0$ for the chirp frequency. In two-photon processes, the energy of the absorbed photon ranges from $3.8c^2$ to $4.0c^2$.
In Fig. \ref{fig4}(d), energy peaks are in the relation of $E=E_i+3\omega_0$ for the fixed frequency, in relations of $E'=E_i+3(\omega_0+bt_1)$ and $E'=E_i+3\omega_0$ for the chirp frequency. Over all, frequency modulation enhances the multiphoton processes by increasing the energy of photon. With the increase of photon energy, the energy spectrum of created electrons becomes wider and higher. Thus, the number of electrons is enhanced. This conclusion is also verified by Fig. \ref{fig3}.

Next, the influence of chirp parameter $b$ on the number of electrons generated under different fundamental frequencies is investigated. In Fig. \ref{fig5}, the final number $N$ of created electrons varies with $b$ for $\omega_0=0.5c^2$, $1.5c^2$, $1.9c^2$, and $2.0c^2$.
For $\omega_0=0.5c^2$, with the increase of chirp parameter $b$, the number of electrons first increases quickly, then decreases slowly.
The number of electrons increases from $1.87$ to $4.76$.
The optimal chirp parameter $b$ is $1.6c^2/t_1$.
For $\omega_0=1.5c^2$, the number of electrons increases from $4.19$ to $5.39$. When $b=0.8c^2/t_1$, the final number reaches the maximum. For $\omega_0=1.9c^2$, the number of electrons reaches the maximum $5.65$ when $b=0.1c^2/t_1$. Overall, with the increase of $\omega_0$, the optimal chirp frequency decreases gradually. The number of electrons is significantly increased by frequency modulation, especially for low fundamental frequencies.
When $\omega_0=2c^2$, with the increase of chirp parameter $b$, the electron number decreases directly. Frequency modulation blocks the creation of electron-positron pairs under high fundamental frequencies.

\begin{table}[htbp]	\label{Table1}
	\centering
	\caption{The maximum, the minimum number of created electrons and the ratio between them for different fundamental frequencies}
    \renewcommand\arraystretch{1.0}
    \renewcommand\tabcolsep{10.0pt}
    \begin{tabular}{l c c c}
          \hline \hline
          $\omega_0(c^2)$ & $N_{min}(b=0)$ & $N_{max}$  & $R(N_{max}/N_{min})$\\
          \hline
          0.1 & 1.87 & 4.58  ($b=1.8 c^2/t_1$) & 2.45 \\
          0.2 & 1.85 & 4.69  ($b=1.7 c^2/t_1$) & 2.54 \\
          0.5 & 1.77 & 4.76  ($b=1.6 c^2/t_1$) & 2.69 \\
          1.0 & 2.30 & 5.13  ($b=1.2 c^2/t_1$) & 2.23 \\
          1.5 & 4.18 & 5.39  ($b=0.8 c^2/t_1$) & 1.29 \\
          1.9 & 5.42 & 5.65  ($b=0.1 c^2/t_1$) & 1.04 \\
          \hline \hline
    \end{tabular}
\end{table}
Table1 I presents the maximum, the minimum number of created electrons and the ratio between them for $\omega_0 <2.0c^2$. When $b=0$,
the number of electrons generated by combined potential wells is the minimum. With the increase of $\omega_0$, the minimum number increases quickly, and the maximum number grows slowly. On the contrary, the optimal chirp parameter $b$ decreases.
Note that the number of electrons reaches the maximum when the sum of $\omega_0$ and $bt_1$ is around $2.0c^2$.
Importantly, from the ratio between the maximum and the minimum, frequency modulation is sensitive to the low fundamental frequencies $\omega_0<1.0c^2$.
For low fundamental frequencies, the maximum number of electrons created by frequency modulation
is comparable to that of high fundamental frequencies.
It means that the number of created electrons at the high fundamental frequencies can be obtained by frequency modulation at the low fundamental frequencies.

In order to explain why the frequency modulation effect is more sensitive to low fundamental frequencies, the number of created electrons varying with the fixed frequency $\omega_0$ is presented in Fig. \ref{fig6}. The number of electrons does not vary monotonously with frequency because the creation mechanism changes as the frequency increases.
In a single subcritical static potential well, the Schwinger mechanism is dominant, and the pair production rate is very small, see Fig. 3(e) in Ref. \cite{Tang2013}.
For an subcritical oscillating potential well, the multiphoton processes are stronger at the higher frequencies, see Figs. 1(a) and (b) in Ref. \cite{Jiang2012}.
In the combined potential wells, the number of electron-positron pairs is the result of the competition of several creation mechanisms, including the Schwinger mechanism, the multiphoton processes, and the dynamically assisted Sauter-Schwinger effect.
In Fig. \ref{fig6}, in the low frequency region, the number of created electrons is relatively small. Although the Schwinger mechanism is strong, the dynamically assisted Sauter-Schwinger effect and the multiphoton processes are both weak.
With the increase of frequency, the Schwinger mechanism weakens. The increase of photon energy results in the enhancement of the multiphoton processes. Among them, the discrete bound states of static potential well also facilitate the multiphoton processes\cite{Tang2013}.
The combination of a high frequency oscillating potential well and a static potential well enhances the dynamically assisted Sauter-Schwinger effect.
Therefore, when the frequency is increased from the low frequency region to $\omega_0=2.0c^2$, there is a significant increase in the number of electrons.
For $\omega_0>2.0c^2$, the ultrahigh frequencies suppress the generation of electrons, resulting in a gradual decrease in the number of electrons. The oscillation of created electron number with frequency is affected by coherence effects in the single-photon regime, see Fig. 3 in Ref. \cite{Jiang2013}.

\begin{figure}[htbp]\suppressfloats
\includegraphics[width=10cm]{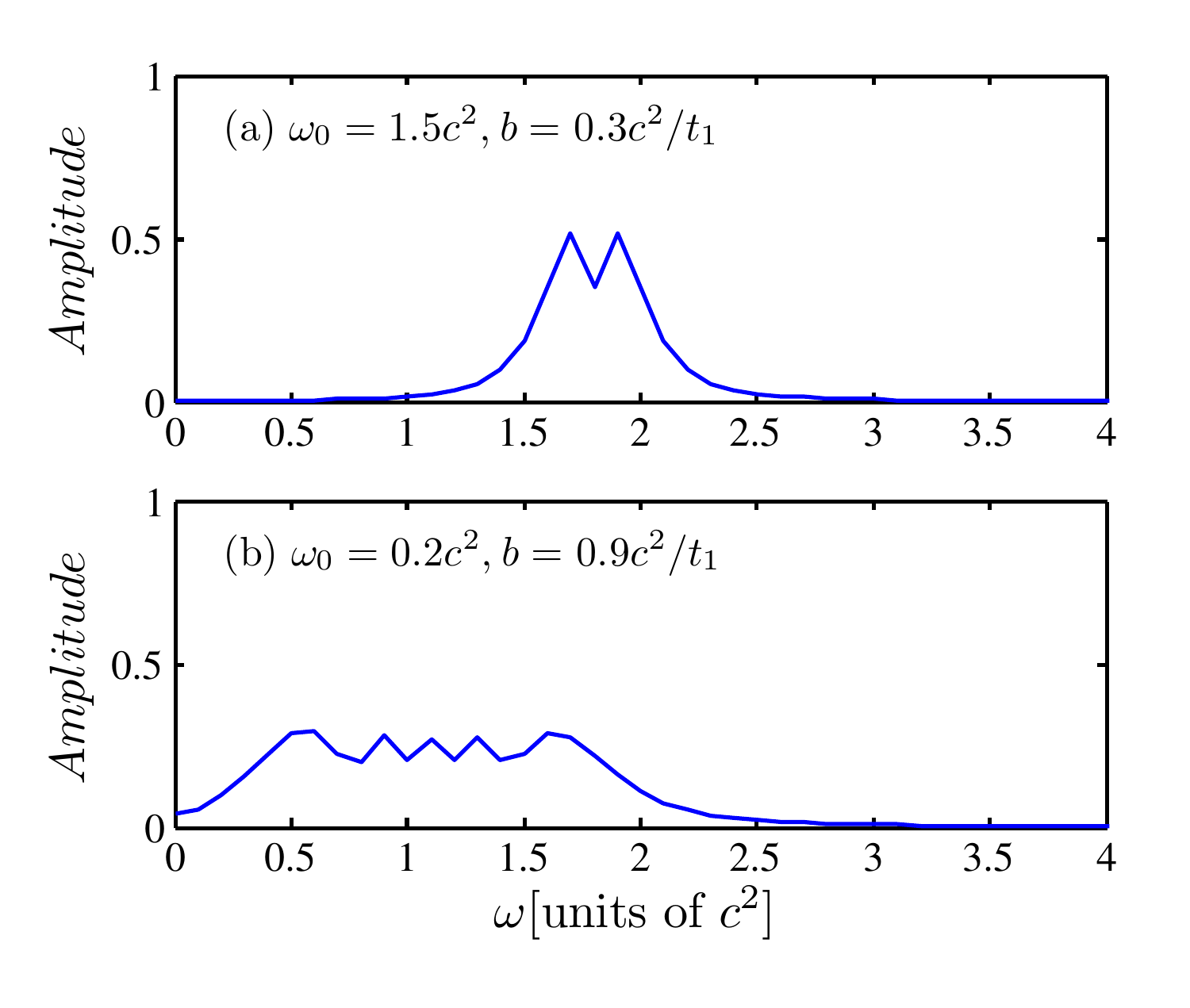}
\caption{\label{fig7} The frequency spectrum of frequency modulated potential well with modulation parameters (a) $\omega_0=1.5c^2$, $b=0.3c^2/t_1$, and (b) $\omega_0=0.2c^2$, $b=0.9c^2/t_1$.}
\end{figure}
The frequency spectrum of frequency modulated potential well is presented in Fig. \ref{fig7}. In Fig. \ref{fig7}(a), modulation parameters are set to $\omega_0=1.5c^2$, $b=0.3c^2/t_1$. The frequency spectrum is symmetric about $\omega_0+bt_1=1.8c^2$. For the high fundamental frequencies, the modulated frequency can be extended to the ultrahigh frequency region by appropriately increasing the chirp parameter $b$. Due to the inhibitory effect of these ultrahigh frequencies, the creation of electron-positron pairs does not benefit from spectrum broadening. As shown in Fig. \ref{fig6}, when the frequency is increased from the high frequency region to the ultrahigh frequency region, the electron number is already very large, so there is not much room for growth. Hence, the enhancement effect of frequency modulation on created electron number is not sensitive to the high fundamental frequencies.
In Fig. \ref{fig7}(b), the spectrum extends from the low frequencies to the high frequencies with modulation parameters $\omega_0=0.2c^2$, $b=0.9c^2/t_1$.
After frequency modulation of the low fundamental frequencies, the weak and high frequency field component appears, which combines with the strong and static electric field to enhance the dynamically assisted Sauter-Schwinger effect. And the enhancement of multiphoton processes after frequency modulation also greatly promotes the generation of electron-positron pairs. In addition, from Fig. \ref{fig6}, there is a lot of room for the increasing of the electron number when the low fixed frequencies are modulated to high frequencies.
Therefore, the enhancement effect of frequency
modulation on created electron number is sensitive to low frequencies.

\begin{figure}[htbp]\suppressfloats
\centering\includegraphics[width=10cm]{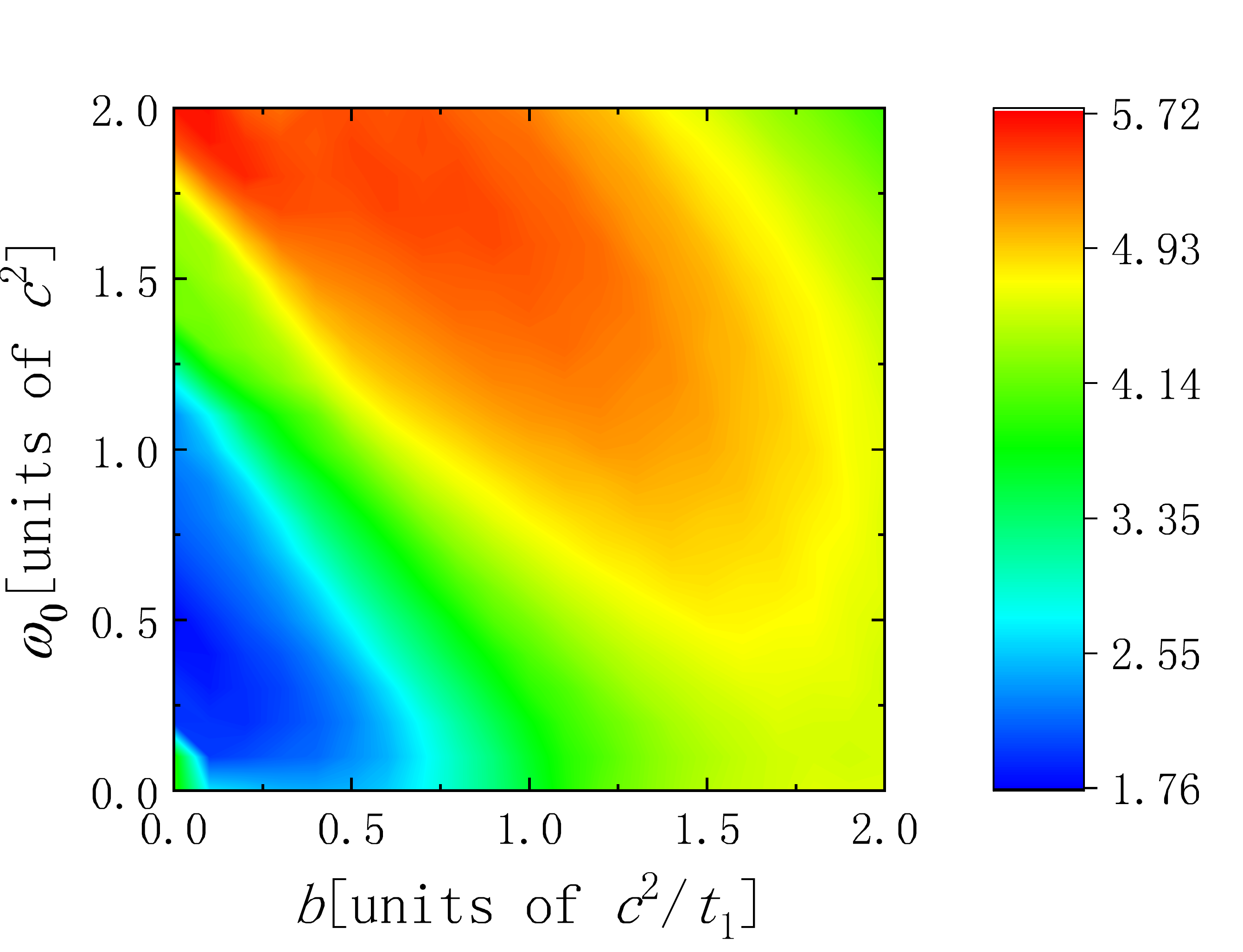}
\caption{\label{fig8} The contour plot of the electron number varying with the fundamental frequency $\omega_0$ and the chirp parameter $b$. Other parameters are the same as in Fig. \ref{fig1}. }
\end{figure}
In Fig. \ref{fig8}, the contour plot of the electron number varying with the fundamental frequency $\omega_0$ and the chirp parameter $b$ is presented. Parameters $\omega_0$ and $b$ are all divided into 20 grids. Discrete parameters $\omega_0$ and $b$ are integer multiples of $0.1c^2$ and $0.1c^2/t_1$, respectively.
In addition, when the sum of $\omega_0+bt_1$ is the same constant, the number of created electrons is roughly the same. With the increase of $\omega_0+bt_1$, the electron number increases rapidly and then decreases slowly. When the fundamental frequency and the chirp parameter are in the relation of $\omega_0+bt_1=2.0c^2$, the number of electrons tends to the maximum, which is consistent with the previous conclusion.
However, there is an anomaly in the extremely low frequency region in the lower left corner of Fig. \ref{fig8}. When $\omega_0+bt_1=0$, the combined potential wells are consisted of two static potential well. The electron-positron pairs are created only by the Schwinger mechanism. The final number of electrons depends on the depth, the width and the edge width of the potential well.
When $\omega_0+bt_1$ increases from $0$ to the extremely low frequency region, the static potential well becomes a slow-oscillating potential well. The number of created electrons is proportional to the effective interaction time, which decreases with the increase of frequency\cite{Wangli2019}.
Therefore, when the potential well changes from static to slow oscillating, the number of created electrons will decrease.
On the whole, frequency modulation improves the creation of electron-positron pairs, especially in low fundamental frequency region.

\section{Conclusion}
The effect of frequency modulation on electron-positron pairs created from vacuum under combined potential wells is investigated by the computational quantum field theory. The combined potential wells are composed of a linearly chirped oscillating potential well and a static potential well. The number and energy spectrum of electrons created under different modulation parameters are studied.
The main results are as follows.

1. The number of created electrons is enhanced significantly by frequency modulation, especially in low fundamental frequency region. By selecting appropriate modulation parameters, the number of created electrons can be increased more than twice.

2. With the increase of the fundamental frequency, the optimal chirp parameter $b$ decreases gradually. The fundamental frequency and optimal chirp parameter roughly satisfy $\omega_0+bt_1=2.0c^2$.

The number of electrons produced mainly by the Schwinger mechanism in low fixed frequencies is small.
With appropriate modulation parameters, the frequency spectrum after linear modulation covers both low and high frequency regions, enhancing the multiphoton processes and the dynamically assisted Sauter-Schwinger effect. The number of electrons increases obviously, and the electron energy spectrum becomes wider and higher. When the frequency spectrum is widened to the ultrahigh frequency region, the creation of electron-positron pairs is inhibited. For high fundamental frequencies, the effective space of frequency modulation is relatively small due to the ultrahigh-frequency suppression effect. So the number of electrons does not increase very much.

In order to investigate the effect of frequency modulation on the multiphoton processes in detail, we use these combination potential wells and find that the number of electrons increases by two times. For a single oscillation potential well or other external fields, the effect of frequency modulation on the yield of electrons may be stronger, which is beyond the scope of present paper.

\begin{acknowledgments}
This work was supported by the Reform and Development Project of Beijing Academy of Science and Technology under Grants No. 13001-2110 and No. 13001-2114; FSZ is supported by the National Natural Science Foundation of China (NSFC) under Grants No. 11635003, No. 11025524, and No. 11161130520; BSX is supported by NSFC No. 11875007; JJL is supported by NSFC No. 12047513.
\end{acknowledgments}

\end{document}